\documentclass[a4paper,12pt]{spieman}  
\usepackage{amsmath,amsfonts,amssymb}
\usepackage{graphicx}
\usepackage{setspace}
\usepackage{gensymb}
\usepackage{lineno}
\usepackage{multirow}
\usepackage{type1ec}
\usepackage{subfigure}
\usepackage{comment}
\usepackage[subfigure]{tocloft}

\title{Key Wavefront Sensors features for Laser-assisted  Tomographic Adaptive Optics Systems on the ELT}

\author[a,b$\dag$]{Thierry Fusco}
\author[c,$\dag$]{Guido Agapito}
\author[b]{Benoit Neichel}
\author[d]{Sylvain Oberti}
\author[b]{Carlos Correia}
\author[d]{Pierre Haguenauer}
\author[c]{C{\'e}dric Plantet}
\author[b]{Felipe Pedreros}
\author[b]{Zibo Ke}
\author[b]{Anne Costille}
\author[b]{Pierre Jouve}
\author[c]{Lorenzo Busoni}
\author[c]{Simone Esposito}

\affil[a]{DOTA, ONERA, Université Paris Saclay (COmUE) [Châtillon]}
\affil[b]{Aix Marseille Univ, CNRS, CNES, LAM, Marseille, France}
\affil[c]{INAF, Arcetri Astrophysical Observatory, Largo E. Fermi 5, 50125-FIRENZE, Italy}
\affil[d]{ESO, Karl-Schwarzschild-Str.2, 85748 Garching b. München, Germany }

\affil[$\dag$]{co-first authors, Send all correspondence to thierry.fusco@onera.fr and guido.agapito@inaf.it }

\cftpagenumbersoff{figure}
\cftpagenumbersoff{table} 

\begin{document} 
\maketitle

\begin{abstract}
Laser Guide Star [LGS] wave-front sensing [LGSWFS] is a key element of tomographic Adaptive Optics system. However, when considering Extremely Large Telescope [ELT] scales, the LGS spot elongation becomes so large that it challenges the standard recipes to design LGSWFS. For classical Shack-Hartmann Wave-Front Sensor [SHWFS], which is the current baseline for all ELT LGS-assisted instruments, a trade-off between the pupil spatial sampling (number of sub-apertures), the sub-aperture field-of-view and the pixel sampling within each sub-aperture is required. For ELT scales, this trade-off is also driven by strong technical constraints, especially  concerning the available detectors and in particular their number of pixels. For SHWFS, a larger field of view per sub-aperture allows mitigating the LGS spot truncation, which represents a severe loss of performance due to measurement biases. For a given number of available detectors pixels, the sub-aperture Field of View [FoV] is competing with the proper sampling of the LGS spots, and/or the total number of sub-apertures. In this paper we propose a sensitivity analysis, and we explore how these parameters impacts the final performance. In particular we introduce the concept of super resolution, which allows to reduce the pupil sampling per WFS, and opens an opportunity to propose potential LGSWFS designs providing the best performance for ELT scales.


\end{abstract}

\section{Introduction}
Europe has just launched the construction of the largest ground-based telescope: the ELT\cite{2020SPIE11445E..1ET}. In operation by 2027, this 40m giant will answer fundamental questions from the search for and characterization of planets to the formation and evolution of the first galaxies of the universe. Adaptive Optics (AO), by correcting in real time aberrations introduced by the atmosphere, is essential to reach the ultimate \textcolor{black}{performance} of this future facility. The ELT has therefore been designed as an adaptive telescope, which will provide images with an angular resolution of less than 10 milli-arcsec in the near infrared. For this purpose, the ELT is equipped with a deformable mirror in its optical train (the 4th mirror of the telescope, alias \textcolor{black}{ELT} M4\cite{2019Msngr.178....3V}), as well as \textcolor{black}{6 to 8} laser stations\cite{2020SPIE11445E..1ET}, in order to create artificial sources (Laser Guide Stars or LGS) for wavefront analysis.
Laser stars have been used in AO on 8/10m telescopes for about \textcolor{black}{twenty} years now, and their \textcolor{black}{performance is} relatively well mastered. However, scaling up to a 40m telescope represents a much bigger challenge than a simple extrapolation of the current solutions, and new concepts are required, especially in the field of wavefront sensing.

In this work we focus on Shack-Hartmann WFS because it is currently the best suited WFS for LGS WFSensing. Although alternative solutions, like Pyramid\cite{Ragazzoni96} and Ingot WFS\cite{2018SPIE10703E..3YR}, exist and are pursued actively \cite{2016SPIE.9909E..6BE,2020SPIE11448E..68A}, they still have to be proved on sky and are therefore deemed not mature enough to be used within the ELT.
Despite being a forced choice for LGS sensing, the SHWFS in the context of ELT has a major downside as it requires a detector with a large number of pixels to satisfy at the same time the requirement on the extended spot sampling, both in term of pixel pitch and overall FoV, and on turbulence spatial sampling\cite{2016SPIE.9909E..5ZG,oberti:hal-02614170,2021A&A...649A.158B}.

Although our work is general, it is particularly interesting for the tomographic adaptive optics systems of a couple of first light instruments of the ELT: MAORY\cite{2020SPIE11448E..0YC} (Multi-conjugate Adaptive Optics RelaY) and HARMONI\cite{2020SPIE11447E..1WT} (High Angular Resolution Monolithic Optical and Near-infrared Integral field spectrograph). We report in the following sections several examples derived from these instruments and the LGSWFS main parameters that we propose can be directly applied to them.

In this paper, we present the challenges of Laser-assisted tomography on ELT (Section \ref{sec:challenges}), and the impact on the design of LGSWFS. Section \ref{sec:tomo&superres} deals with a new concept, dubbed super-resolution, which allows to obtain good tomographic performance, with a reduced number of pupil sampling points. Section \ref{sec:LGSsamp&trunc} deals with the specific application to the ELT case, in particular we analyse the impact of the sub-aperture size, of the spot sampling and truncation and of the noise propagation on the AO system \textcolor{black}{performance}.
Finally, in Section \ref{sec:LGSdesigns} we report potential LGSWFS designs considering the performance sensitivities presented in the previous sections.

\section{Challenges of Laser-assisted tomographic AO on the ELT}\label{sec:challenges}

This section summarizes the main challenges imposes by the use of LGS for an ELT.

\subsection{Tip-Tilt information}
This is a well known problem for AO systems working with LGSs: the image motion (or Tip-Tilt modes) cannot be measured directly from the LGS itself\cite{1992A&A...261..677R}. The transition from 8m to 40m does not significantly change this fundamental limitation, and the LGS AO systems still require the use of a Natural Guide Star (NGS) to get these modes. 
Fortunately, we have some favorable factors for the determination of tip-tilt in the detection of wavefront error: (1) the isoplanatic angle of low order modes like tip-tilt is an order of magnitude larger than that of high order modes. And for an ELT, the outer scale of the turbulence can play a positive role in reducing the overall atmospheric Tip-Tilt energy. (2) The entire pupil of telescope can be used to measure the tip-tilt, and as a result, the NGS used for measuring Tip-Tilt only can be significantly fainter than the usual NGS stars in classical AO. A full ELT pupil should allow to access fainter stars than an 8m telescope, hence potentially increasing the sky coverage. (3) The \textcolor{black}{temporal} bandwidth of the close loop in terms of tip-tilt compensation is about 1/4 of that of the high order \textcolor{black}{(the Zernike polynomial cutoff frequency scales linearly with the radial degree as can be seen in Ref.~\citeonline{Conan1995})}, which can be beneficial for increasing the integration time, hence improving the SNR on faint NGS. On the downside, one has to note that this is true for atmospheric tilt, but telescope vibrations or wind-shake, which becomes much stronger for an ELT, would still require a \textcolor{black}{large temporal bandwidth for measurement and control}.
Note that the NGS performance estimation is provided in a companion paper\cite{10.1117/1.JATIS.8.2.021509}.

\subsection{Cone Effect}

The second main limitation of the laser guide star is the cone effect\cite{1994JOSAA..11..277F}. The laser is focused at finite altitude (somewhere in the middle of the Sodium layer), hence the wave we receive from the laser source is spherical. If an AO system can only use a single LGS beacon to sense the atmospheric turbulence, an error is made; this error is called cone effect or ``Focus Anisoplanatism''. The larger the diameter of the telescope, the more important the cone effect becomes. The mitigation to cone effect is to use more than a single LGS, and perform a tomographic reconstruction of the atmospheric volume above the telescope. This is further detailed in Section \ref{sec:tomo&superres}.

\subsection{Spot elongation}
\label{subsec:spotelong}
The spot elongation comes from the fact that the laser stars are not point objects, but extended sources. Indeed, the layer of Sodium atoms, located at ~90km above the telescope, has a thickness between 10km and 20km. The laser stars resulting from the excitation of these Sodium atoms by the laser light propagated from the telescope have thus a ``cigar'' shape in the Sodium layer. By perspective effect, they appear as extended objects (\textcolor{black}{approximately like} ellipses) on the opposite edge of the telescope pupil. For a 40m telescope, the laser spots have a size between 1arcsecond for those close to the Laser Launch Telescope (LLT) and a maximum expected elongation which can reach up to 25arcseconds. The difficulty is therefore to perform a wave front analysis on highly extended objects, and whose elongation varies in the pupil. There has been several options proposed in the literature to mitigate the spot elongation issue. The proposed solutions include modifications of the laser source itself\cite{Ribak04}, innovative wave-front sensing strategy\cite{2016SPIE.9909E..6LJ} and advanced centroiding algorithms\cite{Gilles06}. Of course these solutions are not exclusive, and a combination of some is possible.\\ The mitigation followed in this work is based on the work developed by Bechet et al. \cite{Bechet2010}. The method consists in weighting the measurements in order to only reject the ones along the long axis of the elongation (truncated), while keeping those along the small axis (not truncated). For each spot, the gradient along the small axis would be kept, while the gradient along the long axis would be rejected if truncated. Thanks to the redundancy of the measurements, and thanks to the fact that in a side-launch configuration the elongated spots of one WFS correspond to the non-elongated spots of another WFS, the expected impact on performance should be small. This is detailed in Section \ref{sec:LGSsamp&trunc}.\\

The above mentioned topic of LGS spot truncation naturally leads to a necessary trade-off between the LGS spot sampling and the total Field-of-View [FoV], \textcolor{black}{in the context of the technological limitation imposed by current state of detectors. In fact, for a fixed number of pixels associated with a SHWFS, and a fixed number of sub-apertures, a potential solution to increase the FoV of each sub-aperture could be to increase the angular size of each pixel.} Usually, the SHWFSs are designed to have a pixel size providing the Shannon sampling of the spots within each sub-aperture \textcolor{black}{(when not photon or read-out noise limited)}. For the LGS object, the smaller spot size that one can expect may be around 1 arcsec FWHM. In this case, the ideal pixel size should be 0.5arcsec. Fixing the pixel size, and if the detector and number of subapertures are fixed, will automatically determine the FoV of the WFS. \textcolor{black}{Increasing the pixel size can then be an option to increase the subaperture FoV and mitigate LGS truncation. However, working with larger pixels introduces an undersampling of the LGS spot, which induces non-linear effects (optical gains) in the centroid measurement \cite{Gratadour2010,Thomas2008,Robertson2017}}. The extreme example would be working with quad-cells (2$\times$2 pixels per subaperture), which are known to be very non-linear \cite{veran2000centroid}. Basically, \textcolor{black}{when working with undersampled spots, the centroiding will become a function of the LGS spot size. For LGSWFS, and because the spot size changes across the pupil, the non-linearity will be different for each sub-aperture, making the wave-front reconstruction very complex. 
This dependence (known as optical gain for the pyramid WFS) has to be calibrated on-line, for example by dithering each LGS with a known signal \cite{vanDam05}. }
Even though this on-line calibration is feasible, it adds complexity to the AO system, and centroid gains remain an additional error in the final performance of multi-LGS systems\cite{neichel2014gemini}. This trade-off has been a key topic in the design and definition of the LGSWFSs for the ELT.

\subsection{Key parameters for Laser WFSensing on the ELT}
As said above, several parameters will be impacting the final performance of the laser assisted AO system for the ELT. In the following we explore the performance sensitivity to several of these parameters. In particular, we investigate the following aspects:
\begin{itemize}
    \item Pupil sampling (a.k.a. number of sub-apertures) and tomographic performance.
    \item LGS spot sampling vs. truncation trade-off.
    \item Noise propagation (flux and RON).
\end{itemize}

Note that in this work we aim at minimizing the sum of all the error terms (fitting, aliasing, noise, linearity and truncation) given the constraints of the available technology in terms of detectors.
As we will show in the following sections, this bring the design in the direction of sub-optimal choice with respect to single issues like linearity and truncation.
In fact, we could design WFSs that completely avoid these last two errors terms, but at the price of a lower overall performance.
In the same way, using a large number of pixels per sub-aperture makes the noise larger, but it is the best overall solution to balance linearity and truncation errors.

Finally, note that in this work we do not focus on temporal error because the WFS characteristics we are interested in have no impact on it and, moreover, its weight in the overall error budget for first generation tomographic AO systems on the ELT is negligible with respect to the other terms presented here.
In fact, it can be easily shown\cite{1993ARA&A..31...13B} that for the median atmospheric condition of Cerro Armazones\cite{2013aoel.confE..89S} we expect a temporal error of about 50nm \textcolor{black}{for a closed loop framerate of 500Hz}, much smaller than the error budget of HARMONI and MAORY (see Ref. \citeonline{Busoni2019}, \citeonline{10.1117/1.JATIS.8.2.021509}).

\section{Tomography and super-resolution}\label{sec:tomo&superres}
Before detailing the optimization of the LGSWFS design for the ELT, we first introduce the concept of super-resolution \cite{ObertiSR,2022Oberti}, and its application for tomographic AO systems. 

\begin{figure}[H]
    \centering
    \includegraphics[width=0.95\columnwidth]{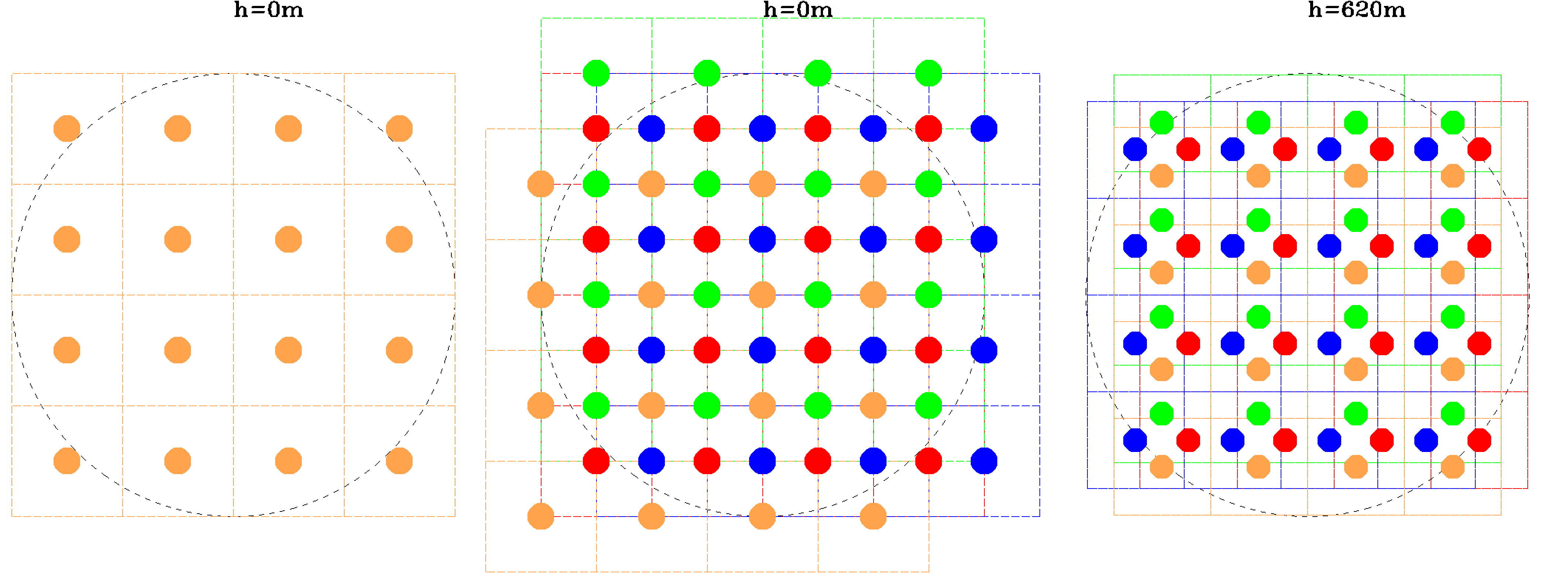}
    \caption{Example of meta-pupil sampling with 4 4$\times$4 LGS WFSs with 50cm sub-apertures and an off-axis angle of 45arcsec. Different color are used to distinguish between the 4 WFSs. Left, altitude is 0m and no super-resolution is present because all sub-apertures are super imposed. Center, altitude is 0m and on purpose mis-aligments give a resolution of half the sub-aperture size. Right, altitude is 620m, the same of \textcolor{black}{ELT} M4, and the sub-aperture shift coming from the 4 different line of sight, $\alpha$=45arcsec, give a naturally induced super-resolution that is approximately half the sub-aperture size.}
    \label{fig:superResSchemes}
\end{figure}

On Wide-Field AO systems [either multi-conjuguate AO (MCAO) or laser tomography AO (LTAO)] several WFSs sense the turbulence volume to get a tomographic information of the disturbances above the telescope. This tomographic information can then be used to optimize the performance in a given direction (single DM $\rightarrow$ LTAO), or across a wider science FoV (several DM $\rightarrow$ MCAO). 
This configuration gives also an additional feature: the different lines of sight of the WFSs increase the number of measurements in the part of the metapupil illuminated by more than one laser \textcolor{black}{beam}. \textcolor{black}{As such, information at an higher spatial frequency than the Nyquist's one given by the number of sub-apertures of a single WFS is accessible \cite{Ellerbroek2001, Wang2012}.} A few examples of the meta-pupil sampling of a tomographic system are shown in Fig. \ref{fig:superResSchemes}. 

\begin{figure}[H]
    \centering
    \includegraphics[width=0.9\columnwidth]{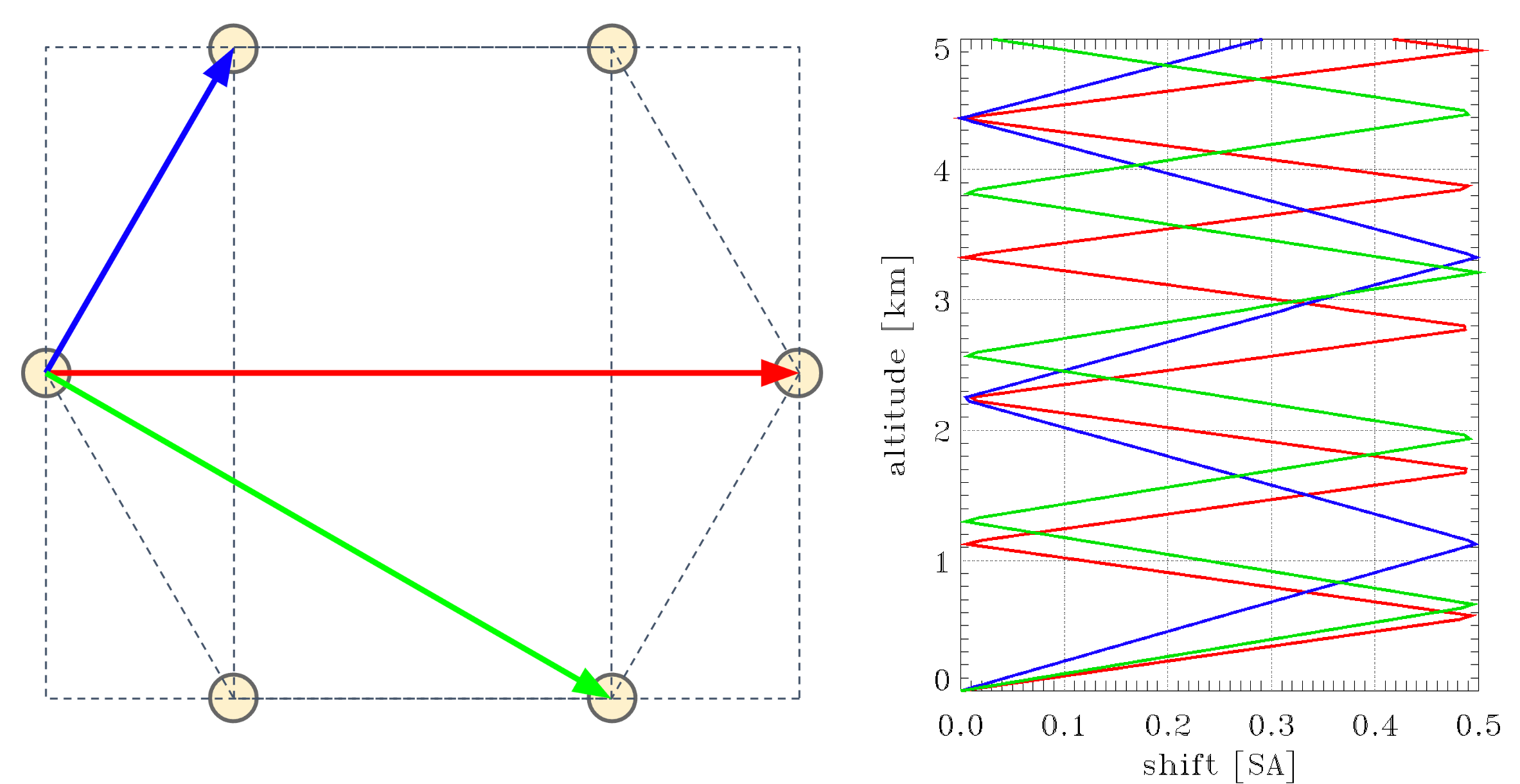}
    \caption{\textcolor{black}{MAORY LGS asterism and the naturally induced relative shift as a function of the altitude (in the range 0-5km) for three couples of WFSs (we used the same colors on the left and right parts of the figure). Sub-aperture (SA in the caption) size is 0.5m and the value reported in the right part of the figure is the results of the modulo 1 operation and values greater than 0.5 are reported as the difference between 1 and the shift value. Where the shift is close to 0.5 sub-aperture the naturally induced super-resolution has the maximum effect and it is completely not effective only when the shifts for all the couples of WFS is 0 (in this range only at an altitude of 0km).}}
    \label{fig:LGSasterismGeometryMAORY}
\end{figure}

 In other words, with multi LGS tomographic AO system \textcolor{black}{without any intentional misalignment,} the geometrical resolution is coming \textcolor{black}{for free, at least} for the turbulent layers in altitude. Unfortunately this free super-resolution is effective in a reduced altitude range because at ground level all WFSs see the pupil with the same geometry and at high altitudes the sensed patch of the WFS separates between each other.
In particular the optimal shift of half sub-aperture is given at an altitude $h_{min}$ considering opposite LGSs and neglecting cone effect:
\begin{equation}
    h_{min}=\frac{d}{4 \alpha} \; ,
\end{equation}
where $d$ is the size of a sub-aperture and $\alpha$ is the LGS asterism radius.
For MAORY, $\alpha$=45arcsec, and HARMONI, $\alpha$=34arcsec, considering a sub-aperture size of 50cm, this value is 573 and 737m respectively, close to \textcolor{black}{ELT} M4 conjugation altitude (620m).
\textcolor{black}{On the other hand}, the altitude at which opposite LGS patch separates at zenith, $h_{max}$, is: 
\begin{equation}
    h_{max}=\frac{Dh_{Na}}{D+ \textcolor{black}{2}\alpha h_{Na}} \; ,
\end{equation}
where $D$ is the telescope diameter and $h_{Na}$ is the average sodium altitude.
\textcolor{black}{This value is 45 and 50km for MAORY and HARMONI respectively ($h_{Na}$=90km).}
This means that a significant part of the turbulence volume is sensed at higher resolution that the one given by the sub-aperture size, \textcolor{black}{ as can be seen in the example of Fig. \ref{fig:LGSasterismGeometryMAORY}. In this figure, the geometry of MAORY LGS asterism and the naturally induced relative shift between couple of WFSs is shown: only at 0km the relative shifts of all the couples of WFS is null, confirming our previous statement (note that this is true also for HARMONI).}

Still the lower part of the ground layer sensing \textcolor{black}{(a few hundreds of m for MAORY and HARMONI)} is limited by the sub-aperture size, unless \textcolor{black}{misalignments} between WFSs sub-aperture grids and pupil is voluntarily introduced. This is the strategy followed, and refered to as Geometrical Super Resolution (GSR).

\begin{figure}
    \centering
    \includegraphics[width=0.95\columnwidth]{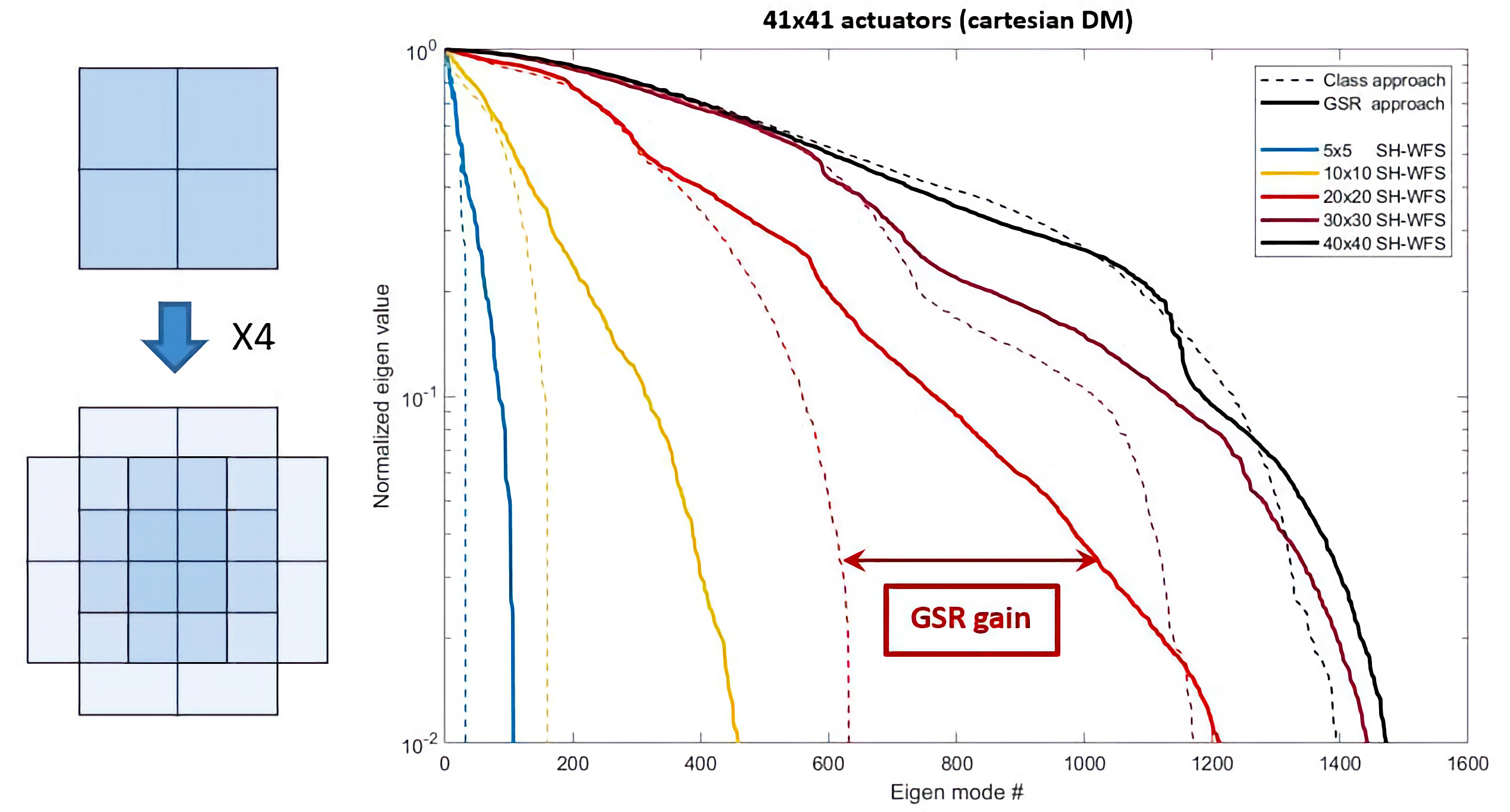}
    \caption{Impact on eigen values of the GSR with respect ot the classical case for configuration with different numbers of sub-apertures. A 41$\times$41 actuator DM is considered here. On the left the effect of GSR for a WFS with 2$\times$2 sub-aperture is shown.}
    \label{fig:ExSR}
\end{figure}

An example of Super Resolution is illustrated in Figure \ref{fig:ExSR} where the eigen value of the interaction matrix of a system composed by a 41$\times$41 actuator Deformable Mirror (DM) \textcolor{black}{conjugated to the pupil} and 4 WFS have been plotted for various cases:
\begin{itemize}
    \item a ``classical case'' for which the 4 WFSs are aligned and see the DM exactly in the same way (it corresponds to a classical single WFS system but we keep the 4 WFSs in order to keep the same number of measurements and normalisation for all the cases). It corresponds to the dashed lines in Figure \ref{fig:ExSR} 
   \item a GSR case for which each WFS is shifted by half a sub-aperture in x and y. It corresponds to the solid lines in Figure \ref{fig:ExSR} 
\end{itemize}
for both the classical and the GSR case we consider a WFS with a various number of sub-apertures (from 5$\times$5 to 40$\times$40, corresponding to various colors in  Figure \ref{fig:ExSR}) knowing that the reconstruction basis remains the same (the 41$\times$41 DM actuator basis for all the cases). 

 \textcolor{black}{Figure \ref{fig:ExSR} clearly shows the gain brought by the GSR when compared to a classical approach, in terms of number of accessible modes.} In particular it is shown that we have access to typically the same number of modes with a 30$\times$30 WFS with GSR \textcolor{black}{as} with a 40$\times$40 with a classical approach. Even though the eigen values are slightly lower for 30$\times$30 with GSR than 40$\times$40 with classical approach, the quadratic sum of the eigen value only show a 15\%  decrease and therefore should not bring any significant degradation in terms of noise propagation.



\begin{figure}[H]
    \centering
    \includegraphics[width=0.6\columnwidth]{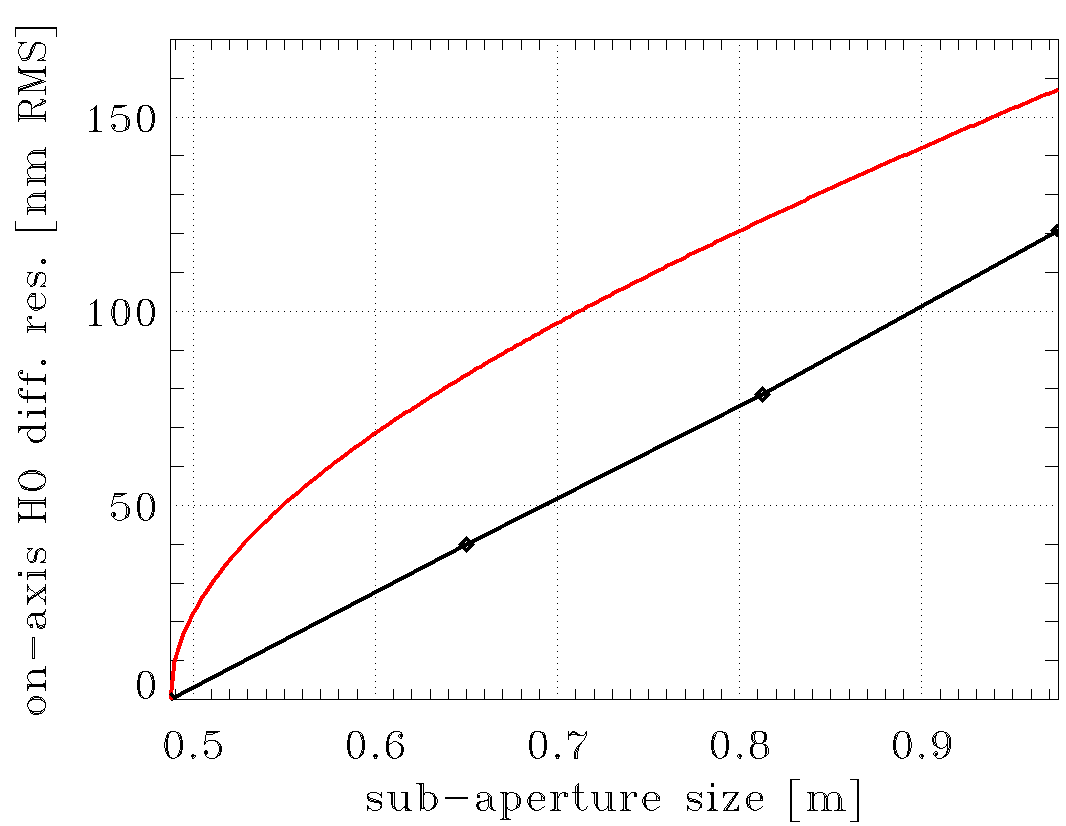}
    \caption{On-axis MAORY High Order (excluding tilts) differential (with respect to a sub-aperture size of 0.4875m) residual as a function of sub-aperture size. High flux, no elongation. Red line, theoretical residual considering fitting error increase due to lower sampling. Black line is lower thanks to naturally induced super-resolution \textcolor{black}{(that means without any intentional misalignment)}.}
    \label{fig:MAORYonaxisHOratioOnaxisSAsize}
\end{figure}

\begin{figure}[H]
    \centering
    \includegraphics[width=0.99\columnwidth]{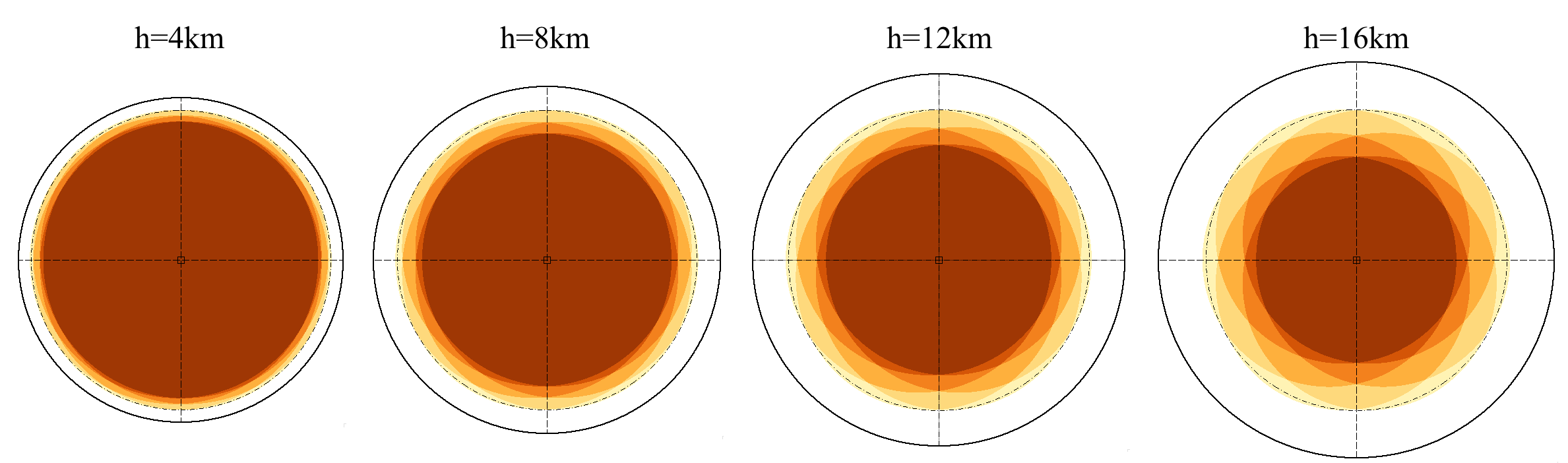}
    \caption{LGS beams overlap at different altitudes for MAORY working at a zenith angle of 30deg. Decreasing color luminance is used to show an increasing number of overlapping beams. Dotted circles correspond to the pupil size, 39m, while white portion corresponds to the area where the FoV is not sensed and it mainly corresponds to technical FoV.}
    \label{fig:MAORYbeams}
\end{figure}

The application of \textcolor{black}{naturally induced} super-resolution for MAORY \textcolor{black}{(that means
without any intentional misalignment)} is shown in Fig. \ref{fig:MAORYonaxisHOratioOnaxisSAsize}.
Here it is interesting to note that the \textcolor{black}{expected} error increase \textcolor{black}{given by} fitting error, shown in red, is following the law $k_f \sqrt{d_{}^{5/3}-d_0^{5/3}}$, while the one found for super-resolution is following the law $k_s(d-d_0)$, where $k_f$=194, $k_s$=248 and $d_0$=0.4875.
We verified that all the tested configurations are able to properly correct the turbulence with all degrees of freedom of \textcolor{black}{ELT} M4 so the error reported above is \textcolor{black}{not due to the loss of sensitivity to the highest spatial frequencies of the DM.}
Moreover working in high flux conditions we consider that measurement noise error is negligible.
Hence, we hypothesize that the difference in performance is mainly due to \textcolor{black}{the ground layer sensing (see Fig.\ref{fig:LGSasterismGeometryMAORY}),} spatial aliasing and tomographic error.
In particular for the tomographic error, super-resolution is valid only where the LGS beams are super-imposed, \textcolor{black}{and it has a lower efficiency where the overlap of the beams is partial. Note that super-resolution is completely useless where the turbulence is sensed by single beams. So the super-resolution is typically less effective on the external part of the meta-pupil at the highest altitudes, where, as can be seen in Fig. \ref{fig:MAORYbeams}, the beams are not super-imposed.}

%

%
%

\begin{figure}[H]
    \centering
    \includegraphics[width=1\columnwidth]{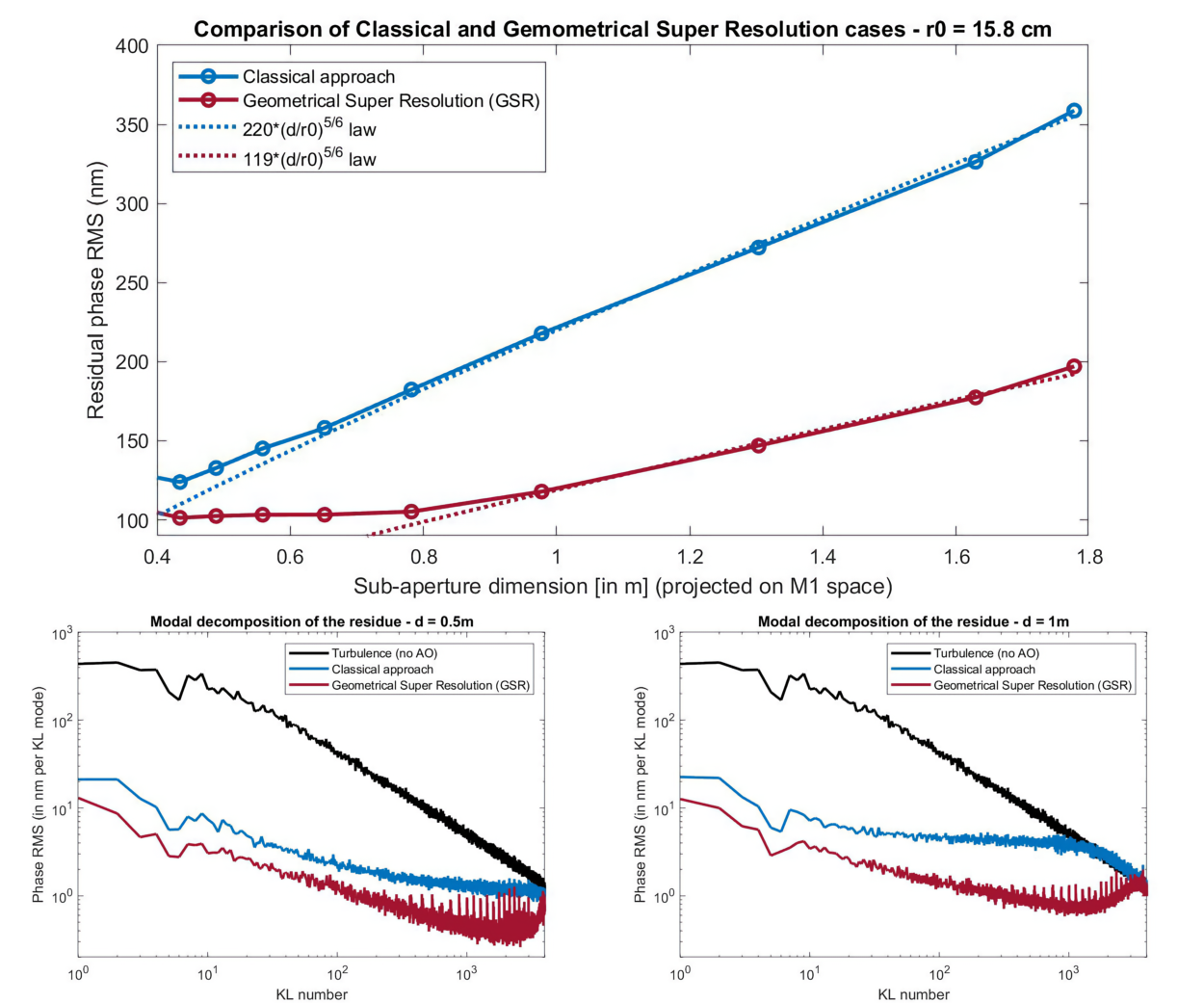} 
    \caption{Impact on performance of sub-aperture size. \textcolor{black}{Note that all the turbulence is located at ground level (0m) as well as the DM (so no natural induced super-resolution is present)}. [Top] residual phase RMS as a function of the sub-aperture dimension - [Bottom] turbulence and residual variance modal decomposition on a Karhunen-Lo{\'e}ve base for different configurations for respectively 0.5m (80 sub-apertures in the diameter) and 1m (40 sub-apertures in the diameter) for each plot, [blue line] stands for classical AO, [red line] stands for Geometrical Super Resolution (GSR). [Black line] stands for the full turbulent (no AO) case.}
    \label{fig:super_res_vs_sspup}
\end{figure}

The results for the HARMONI case are illustrated in Fig. \ref{fig:super_res_vs_sspup}.
\textcolor{black}{Here all the turbulence is located at ground level (0m) as well as the DM, so we can focus only on the the GSR. In fact, the system does not benefit from naturally induced super-resolution in such configuration (therefore these results are also valid for MAORY).}
%
%
The residual phase of the ``classical appraoch'' increases with a slope equal to $(d/r_0)^{5/6}$ ($d$ is the sub-aperture size)
\textcolor{black}{in a similar fashion as the fitting error in a SCAO system.}
Instead, when super-resolution is introduced, following the features listed \textcolor{black}{above in this section}, 
the slope decreases until it becomes almost flat in the range 100-60 sub-apertures ($d$=0.400-0.667m).

\section{LGS sampling vs. spot truncation}\label{sec:LGSsamp&trunc}

\subsection{LGS spot size and sampling}
The sampling impacts both the LGSWFS linearity and noise propagation. For the former, as soon as one considers a sampling lower than Shannon (2 pixels per FWHM), an optical gain will appear. For the LGS, this optical gain (conversion between CoG in pixel, and CoG in arcsec) will be different for X and Y, and different for each subapertures. This gain changes with seeing, and monitoring it requires complex calibration procedures. The extreme example are WFSs working with quadcells, and it has been shown that in this case it could represent an important performance limiting factor (e.g. GeMS). But Shannon sampling may become very expensive in terms of pixels required by the LGWFS, and undersampling the LGS spots may be required. The impact of LGS spot undersampling on the centroiding performance has been studied in previous papers (Gratadour et al.\cite{2010SPIE.7736E..1AG}, Nicolle et al.\cite{2004OptL...29.2743N} and Ke, Pedreros et al. \cite{10.1117/1.JATIS.8.2.021511}), and is not detailed here. The conclusion (coherent between all the papers) is that one can go down to 1 pixel per FWHM while keeping the final error to a reasonable level. 


\subsection{LGS spot truncation}
The second big problem when dealing with LGS for an ELT comes from the spot elongation. Spot elongation per se may not be a problem, as long as enough photons are available to perform centroiding. The main issue comes from spot truncation, as this introduces biases in the tomographic reconstruction. Biases may be evolving quickly, at a rate following the Sodium layer spatial evolution. So truncation and associated biases should be minimized as much as possible. This is made possible by (i) maximising the FoV per sub-aperture and (ii) using a regularized reconstruction.
\begin{figure}[H]
    \centering
    \includegraphics[width=1\columnwidth]{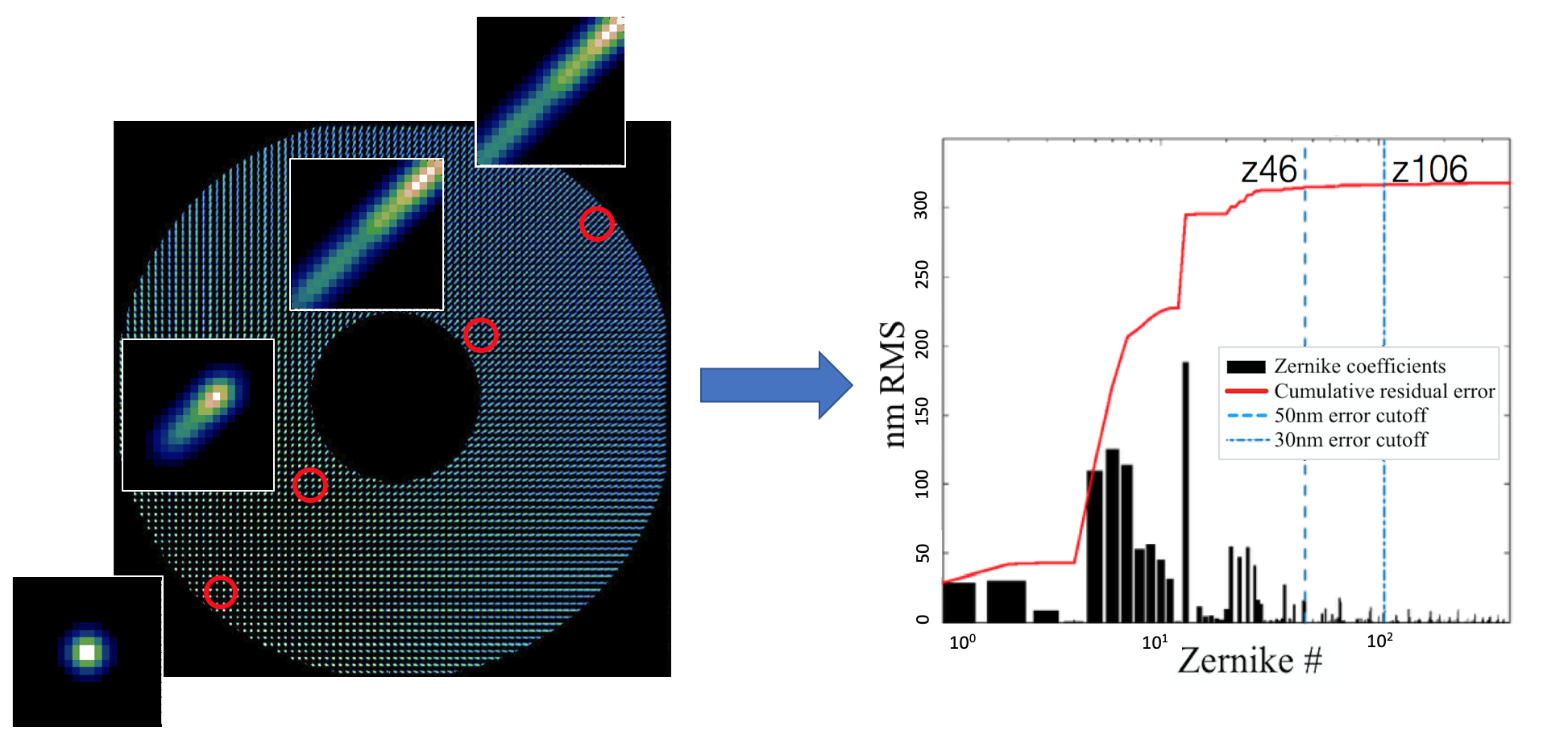}
    \caption{Example of spot truncation in the extreme case of a very extended sodium profile and a small ( 10"$\times$10") FoV WFS. [Left] One WFS image, [Right] the aberrations produced by the spot truncation on the 6 LGS after propagation through the tomographic reconstructor}
    \label{fig:spottrunc}
\end{figure}
Of course these aberrations are statics for a given profile but sodium profiles can evolve quite rapidly (with timescale of a few seconds typically) and more important, when mixed with turbulence these effects are randomized and become quite difficult to handle with a dedicated truth sensor. The best solutions from a system point of view are therefore twofold:
\begin{itemize}
    \item to increase the Shack-Hartman sub-aperture FoV in order to reduce the truncation effect at its root
    \item add a numerical over-regularisation process
    \cite{fusco:hal-02863679,oberti:hal-02614170}
    to over-penalize the signal coming from the most truncated in the reconstruction process 
\end{itemize}

\begin{figure}[H]
   \centering
    \includegraphics[width=0.7\columnwidth]{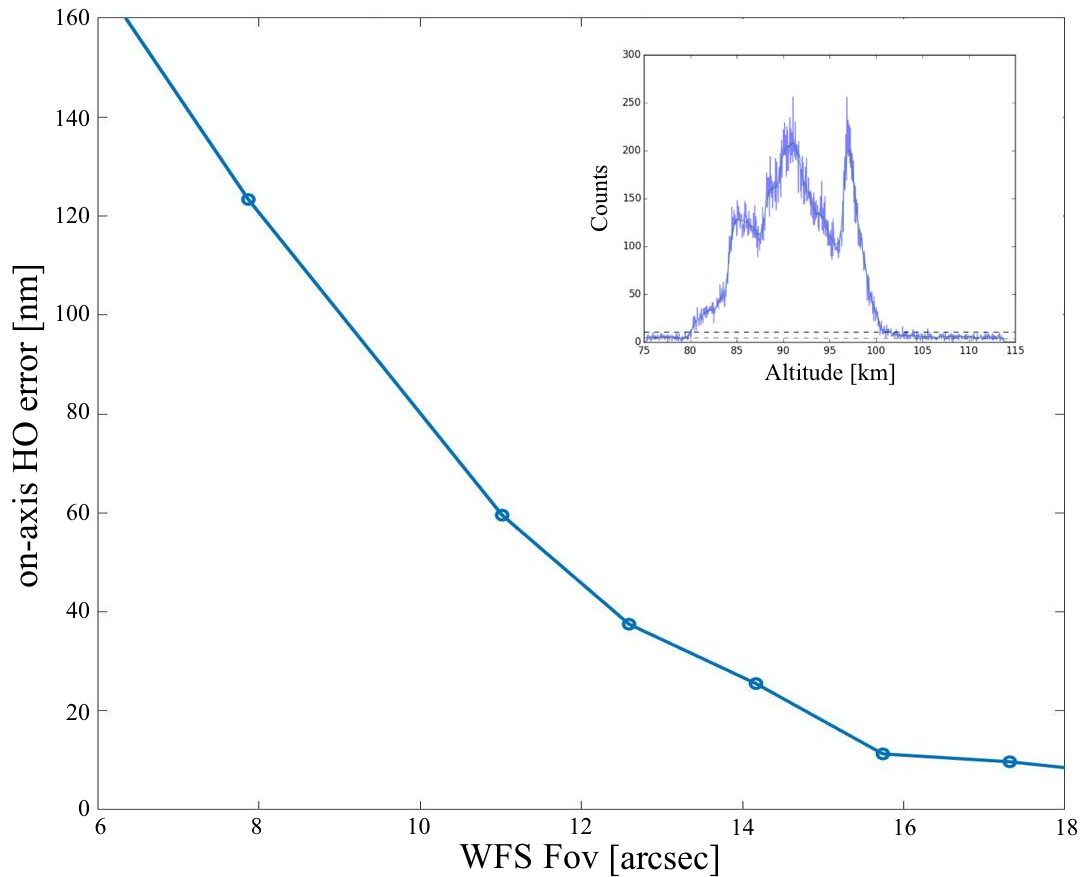}
    \caption{On-axis High Order (excluding tilts) differential (with respect to a FoV of 22 arcsec) residual as a function of WFS FoV. 68$\times$68 sub-apertures ``Very-wide'' sodium profile\cite{2014A&A...565A.102P}. Sodium profile structure is given at Zenith, the simulation have been made for a 30° zenith angle with the according stretch of the profile altitude. For this particular profile the width is estimated to 22km \textcolor{black}{(this gives on the WFS spots with long axis FWHM up to 25"). Note that this error is field dependent, but the on-axis trend is representative of the overall error in the corrected FoV.}}
    \label{fig:MAORYonaxisHOerrorWFSfov}
\end{figure}

We have evaluated the impact of the sub-aperture FoV, and spot truncation on performance in Figure \ref{fig:MAORYonaxisHOerrorWFSfov}. As can be seen from Figure \ref{fig:MAORYonaxisHOerrorWFSfov}, the truncation error can become very significant when the FoV decreases down to 10arcsec. A reasonable FoV per sub-aperture would be to accomodate at least 15 to 16arcsec. 

%

\section{Noise propagation}\label{sec:noiseProp}
The last parameter that must be taken into account for the design of the LGSWFSs for the ELT is the sensitivity to noise, both photon and detector noises.
We recall that for classical AO system, the final noise contribution : $\sigma^2_{noise}$ can be approximated by the following equation\cite{RoddierBook1999}:

\begin{equation} 
\sigma^2_{noise,i} = \sigma^2_{WFS,\lambda_{wfs}} \left(\frac{\lambda_{wfs}}{\lambda_{im}}\right)^2*P_{rec,i}*T_{filtering}  \text{   \textcolor{black}{[in rad² at $\lambda_{wfs}$]}} 
\end{equation}
And the propagated noise by:
\begin{equation} 
\sigma^2_{noise,prop} = \sum_{i=1}^{n_{corr}}\sigma^2_{noise,i} \text{   \textcolor{black}{[in rad² at $\lambda_{wfs}$]}}
\end{equation}
where $\sigma^2_{WFS,\lambda_{wfs}}$ is the variance of the noise at the WFS level, $\lambda_{wfs}$ is the WFSensing wavelength, $\lambda_{im}$ is the imaging wavelength, $P_{rec,i}$ is the coefficient that takes into account the propagation of noise in the phase reconstruction process and $T_{filtering}$  is the coefficient that takes into account the propagation of noise in the temporal filtering process.
If we assume that statistically all the WFS sub-apertures have the same noise propagation, then we can write:
\begin{equation} 
\sigma^2_{noise,prop} =  \sigma^2_{WFS,\lambda_{wfs}}*\left(\frac{\lambda_{wfs}}{\lambda_{im}}\right)^2*T_{filtering}\sum_{i=1}^{n_{corr}}P_{rec,i} \text{   \textcolor{black}{[in rad² at $\lambda_{wfs}$]}} 
\end{equation}

$T_{filtering}$ mainly depend on the temporal aspect of the AO loop: integration time, read-out time, control law characteristics and loop delay. For a basic (and yet widely used) integrator-like scheme\cite{RoddierBook1999}, assuming a 2 frames delay and a loop gain of 0.5 gives $T_{filtering} \textcolor{black}{=} 1/10 $.

The main differences between Laser tomographic and classical AO are the following: 
\begin{itemize}
    \item $\sigma^2_{WFS,\lambda_{wfs}}$ is no more the same for all the sub-aperture and will depend on the spot elongation.
    \item $P_{rec,i}$ will include a tomographic reconstruction and will differ from a classical modal basis noise propagation matrix as described in Ref. \citeonline{1992A&A...261..677R} for example.
\end{itemize}

An interesting case to study here is the following: a $N_{lgs}$ LGS system, without elongation, associated to a tomographic case where all the turbulence will be located in the telescope pupil. In that particular case, $P_{rec LTAO,i} = \frac{1}{N_{lgs}} P_{rec, SCAO,i}$ and $\sigma^2_{WFS,\lambda_{wfs}}$ is the same for all the sub-apertures of all the LGS. The tomographic noise propagation is therefore nothing but $\frac{1}{N_{lgs}}$ the SCAO noise propagation 

This extreme case allows us to draw the ultimate limits in terms of performance for any wave-front sensing configurations. It will also help to understand the influence of the critical parameters of the WFS design to be adjusted in final design choice.  

Let us first study the impact of the main WFS parameters in the ultimate case of photon noise limited detector. For that we study the influence of two main parameters: the spot FWHM (\textcolor{black}{mainly produced by the combination of the spatial extension of the Laser and its propagation through the atmosphere}) and the pixel size (projected on sky). 

For the photon noise limited case, considering a simple centre of Gravity [CoG], the analytical expression of $\sigma^2_{WFS,\lambda_{wfs}}$ \textcolor{black}{can be derived from the standard expressions \cite{Rousset99, 2004OptL...29.2743N}, and expressed as follow:} 
\begin{equation}
\sigma^2_{ph,\lambda_{wfs}} = \frac{\pi^2}{2\ln(2)} \frac{1}{N_{ph}} \left(\frac{\sqrt{\text{FWHM}_{spot}^2 + \text{FWHM}_{pixel}^2}}{\text{FWHM}_{diff}}\right)^2 \text{   [in rad² at $\lambda_{wfs}$]} \label{eq:pixscale}
\end{equation}
where $\text{FWHM}_{diff}^2$ is the diffraction limited size of the sub-aperture, $\text{FWHM}_{spot}$ the FWHM of the spot size in the sub-aperture focal plane and $\text{FWHM}_{pixel}$ the pixel response. Note that for this last we have assumed that the pixel FWHM is equal to the physical size of the pixel projected on sky. This is not always the case but it remains a fairly good approximation for classical detectors used in AO.  
More importantly, the previous equation is highlighting two fundamental points for a SH :
\begin{itemize}
    \item the pixel size of the WFS has to be smaller than the spot size in order to have a negligible impact on the noise propagation;
    \item we can defined a ``Number of useful photon'' $N_{usefull}$ as 
    \begin{equation}
    N_{usefull} = N_{ph} \left(\frac{\text{FWHM}_{diff}}{\sqrt{\text{FWHM}_{spot}^2 + \text{FWHM}_{pixel}^2}}\right)^2 
\end{equation}
in other words, enlarging the spot or working with large pixel is equivalent to reducing the WFS efficiency. As an example, working with 1arcsec pixel combined with 1" FWHM spot is 2.25 times more efficient \textcolor{black}{(in terms of noise variance)} than a 1.5" pixel and 1.5" FWHM spot.
\end{itemize}

In the case of uniform Gaussian noise induced by the detector read out (a.k.a. RON), the number of pixels becomes a critical parameter. Again, considering a classical CoG, the analytical expression of the measurement variance is 
\begin{equation}
\sigma^2_{ron,\lambda_{wfs}} = \frac{\pi^2}{3} \left(\frac{\text{FWHM}_{pixel}}{\text{FWHM}_{diff}}\right)^2\left(\frac{ron}{N_{ph}}\right)^2 \left(\frac{N_s}{N_d}\right)^2 \text{   [in rad² at $\lambda_{wfs}$]} 
\end{equation}
where $\frac{\text{FWHM}_{pixel}}{\text{FWHM}_{diff}}$ converts pixel error in radian. $ron$ stands for the read-out noise, $N_s$ stands for the total number of pixel involved in the CoG computation and $N_d$ the number of pixel in the spot FWHM. 
\textcolor{black}{Assuming that $N_s$ is computed as the ratio of a spot \textcolor{black}{area} equal to 2 times the spot FWHM and the pixel size, then $\frac{N_s}{N_d} = 4*\frac{\text{FWHM}_{spot}}{\text{FWHM}_{pixel}}$ as soon as  $\text{FWHM}_{pixel} < \text{FWHM}_{spot}$ and   $\frac{N_s}{N_d} = 4$ otherwise. }
\begin{equation}
 \sigma^2_{ron,\lambda_{wfs}} = \frac{\pi^2}{3} \left(\frac{\text{FWHM}_{pixel}}{\text{FWHM}_{diff}}\right)^2\left(\frac{ron}{N_{ph}}\right)^2 \left(4*\frac{\text{FWHM}_{spot}}{\text{FWHM}_{pixel}}\right)^2 \text{   [in rad² at $\lambda_{wfs}$]} 
\end{equation}
And finally the ron contribution to WFS measurement variance can be approximated by 
\begin{eqnarray}
 \sigma^2_{ron,\lambda_{wfs}} & =\frac{16\pi^2}{3} \left(\frac{ron}{N_{ph}}\right)^2  \left(\frac{\text{FWHM}_{spot}}{\text{FWHM}_{diff}}\right)^2 & \text{if $\text{FWHM}_{spot} > \text{FWHM}_{pixel}$} \\
                        & =\frac{16\pi^2}{3} \left(\frac{ron}{N_{ph}}\right)^2 \left(\frac{\text{FWHM}_{pixel}}{\text{FWHM}_{diff}}\right)^2
                        & \text{if $\text{FWHM}_{spot} < \text{FWHM}_{pixel}$}
\end{eqnarray} 
Here again it is very easy to understand the importance of the spot FWHM and of the pixel size in the noise produced by any SH measurement. It is essential to minimize these two parameters in order to have the smallest possible noise and keep the pixel size close (but slightly smaller) than the spot FWHM.



\begin{figure}[H]
   \centering
    \includegraphics[width=0.9\columnwidth]{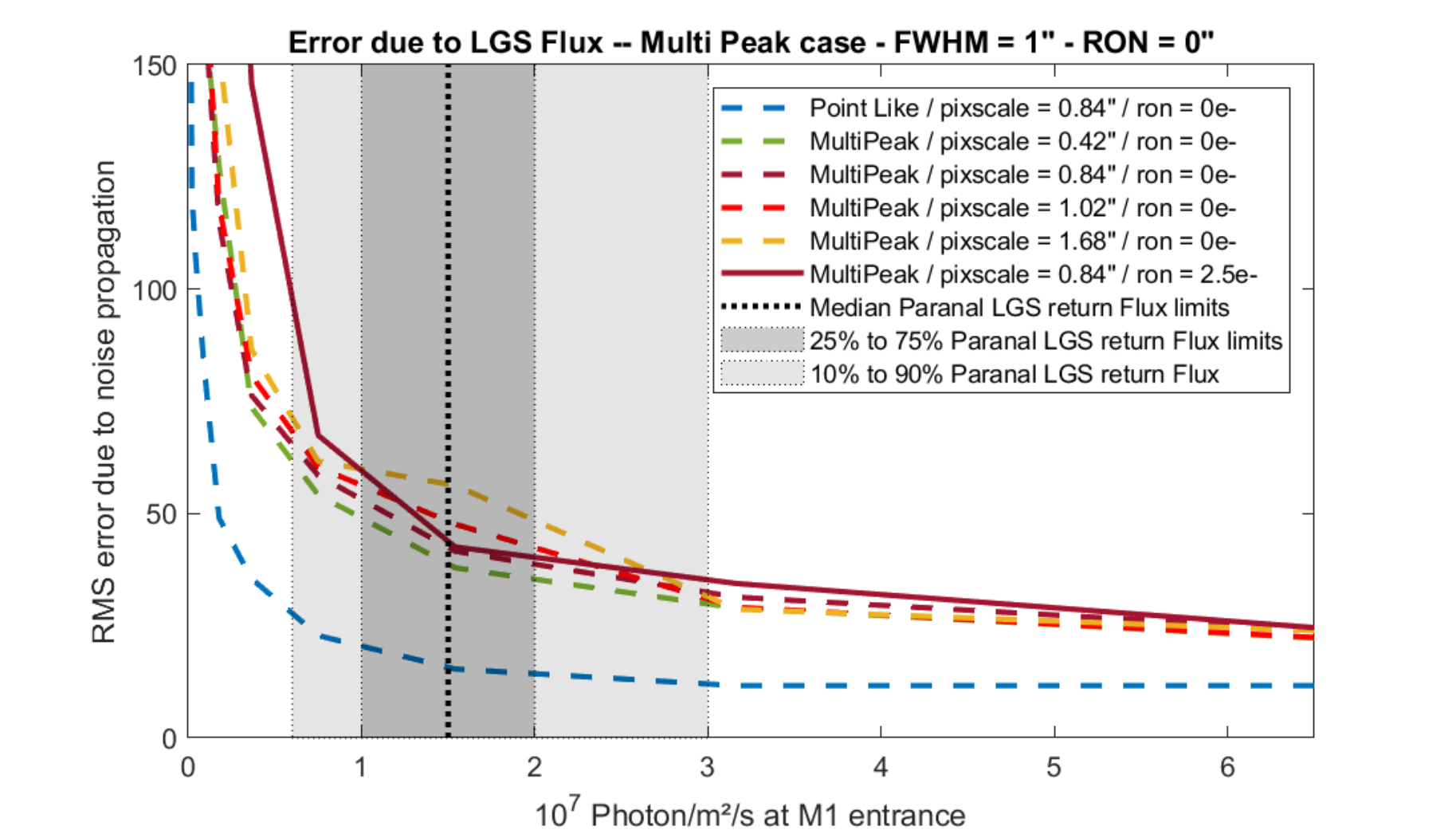}
    \caption{Impact of sodium return flux on performance. Error RMS as a function of the sodium return flux for point like and ``multi-peak''\cite{2014A&A...565A.102P} sodium profile, different pixel scales \textcolor{black}{ and 68$\times$68 sub-apertures.}}
    \label{fig:noise}
\end{figure}

%
\textcolor{black}{Although it is quite easy to develop analytical formulae under the hypothesis of a single WFS and symmetric and identical (statistically speaking) spot shape for all the sub-aperture, in the case of a tomographic AO system none of these hypothesis remains true.} If the tendencies highlighted by the theoretical developments remain correct, it is important to have a full simulation in order to have a better understanding of the limitations and behaviour in a full LTAO \textcolor{black}{or MCAO} case.
Using full end-to-end simulation but looking at the noise propagation term only,  we have estimated the performance sensitivity to photon and RON noise. This is illustrated by Figure \ref{fig:noise}.
Several conclusions can be derived from this figure:
\begin{itemize}
\item for ELT \textcolor{black}{applications} and scaling flux returns measured after more than 4 years of LGS operation at Paranal with AOF, the noise error terms for extended spot (we have considered here some extreme case of Sodium profile) remains quite reasonable: around 50nm rms.
\item The very first increase of noise \textcolor{black}{from blue to green line} is due to the spot elongation. For photon noise only, going from a Point link source to an extended LGS increases the noise from 15 to 41 nm rms, that is an quadratic increase of 38 nm rms.
\item The pixel size (as shown in Eq \ref{eq:pixscale}) is also an important parameter. In the photon limited case its increase by a factor 4 (from 0.42 to 1.68") has led additional noise error of 40 nm rms (quadratic increase). In other word a factor 4 of sampling increase has the same impact \textcolor{black}{on noise propagation} than the spot elongation itself. 
\item \textcolor{black}{Note that for the specific case of a pixelscale of 1.68", non-linearities (pixel scale larger than spot FWHM) should have started to degrade the performance. These non-linearities are not properly modeled in the simulations, and these results are then optimistic. For a real system to work with such a large pixel scale, on-line gain tracking would be required (see Section \ref{subsec:spotelong}).} 
\item Adding Read-Out-Noise has a dramatic impact for low flux cases. In the typical range of return flux we will have to consider with the ELT and considering RON smaller than 3e-, the impact remains very small. Note that we have considered here a very basic 3$\sigma$ threshold algorithm for the centroiding measurements. A more clever approach (Weighted Center of Gravity, Correlation or Match Filter) will further reduced the impact.  
\end{itemize}

In conclusion, with an end-to-end simulation and looking at the noise propagation error term only, we have shown that noise propagation error remains quite low for a LTAO \textcolor{black}{or a MCAO} system working with Paranal typical flux return. \textcolor{black}{The RON is not an major contributor for the level considered here (2.5 e-)}. Pixel scales have to remain reasonably small with a good (not to say optimal) compromise around the typical size of the spot small axis FWHM (that is around 1").   
%

\section{Potential LGSWFS designs}\label{sec:LGSdesigns}
In the previous sections we have derived the performance sensitivity to several factors, like the pupil spatial sampling, LGS spot sampling and impact of truncation. In this section, we have now to put these sensitivity analyses in front of expected LGS parameters, and eventually provide recommendations on what a LGSWFS \textcolor{black}{could} be for the ELT. This should not be seen as a design definition, but guidelines for tomographic AO systems for the ELT. Of course, LGSWFS design choices strongly depends on the Sodium characteristics, so we begin this section by an overview of typical Sodium  parameter distribution derived from the VLT operation.

\subsection{Sodium typical characteristics}
The Adaptive Optics Facility (AOF\cite{2010Msngr.142...12A}) has been operational on the UT4 telescope at Paranal since 2017, providing access to a deformable secondary mirror (DSM) and four LGS units (4LGSF) for a variety of instruments. The GALACSI AO module is also part of the modules delivered within the AOF, bringing AO correction capabilities for the MUSE (Multi Unit Spectroscopic Explorer) instrument, and is also a great tool to gather statistics on the Na layer characteristics thanks to the continuous logging of the AO telemetry data. 

The LGS spot size on sky is dependent \textcolor{black}{on} the seeing but also on the laser launch telescope design.
The on-sky \textcolor{black}{spots measured} during the \textcolor{black}{4LGSF} commissioning phase \textcolor{black}{had a} FWHM of 1.05" for the small axis and 1.61" for the long axis, with the atmosphere seeing being 0.6" and UT4 pointing altitude of 60 degrees. 
\textcolor{black}{The difference between the small and long axes comes from the spot elongation, due to the extended structure of the Na layer, the position of the launch telescope on the side of the pupil, and the distance between the launch telescope and the position in the pupil where it is measured. On-sky measurements were also performed} at different pointing altitudes showing no significant dependence, and with different seeing conditions. With a seeing of 1", the on-sky measured spot size was about 1.35" for the small axis and 2.1" for the long axis. These spot sizes were measured on the UT4 guiding camera for the full 8-m aperture and with the LGS in the center of the field. \textcolor{black}{The launch telescopes for the LGS on the ELT will be identical to the ones of the VLT, and the spot size validated on-sky can thus be used for ELT AO simulations.}

\begin{figure}[H]
    \centering
    \includegraphics[width=0.5\columnwidth]{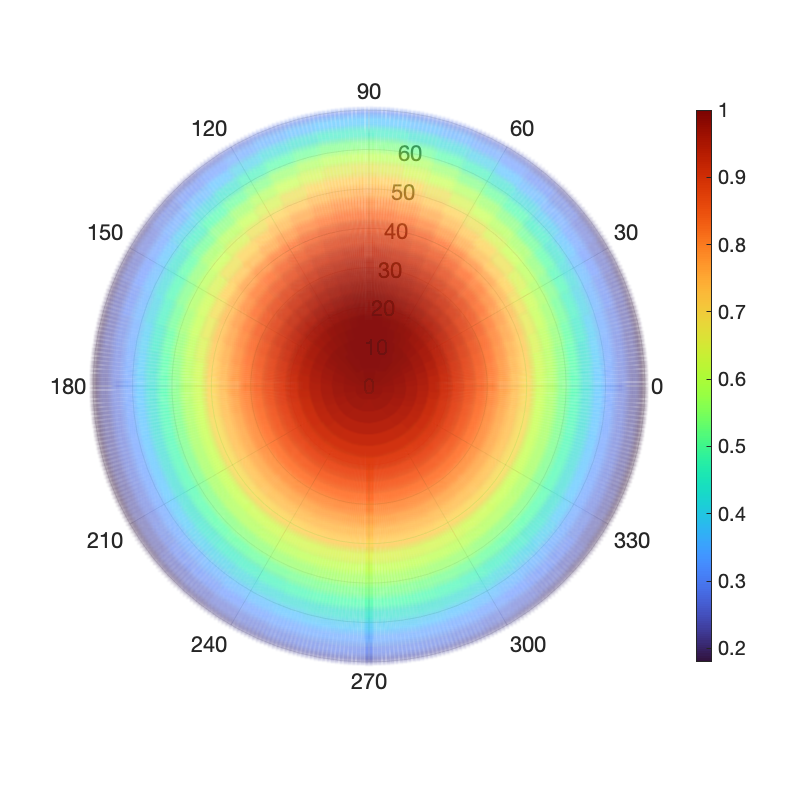}
    \caption{Sky plot of the theoretical LGS flux return at Paranal for a seeing of 0.8”, normalized to the maximum value, based on the LGSBloch model (N=90, E=0). \textcolor{black}{The variations are mostly dependent, beside the pointing direction, on the local direction and magnitude of the geomagnetic field, and on the laser power, polarisation state, and re-pumping efficiency}.}
    \label{fig:LGSFluxReturnVsAltAzNormalized}
\end{figure}

During on-sky operation of GALACSI, the fluxes of all four WFSs are logged regularly. Data are available since July 2017 and are processed to derive the statistics of returned flux, converted to photons per second and per m$^2$ at M1 entrance. The first parameter impacting the flux return is the pointing direction of the telescope at the time of the observation\cite{2010SPIE.7736E..0VH}. Fig.\ref{fig:LGSFluxReturnVsAltAzNormalized} presents the sky plot of the simulated LGS flux return for a seeing of 0.8”, a sodium abundance of 4x10$^{13}$ atoms/m$^2$, and a 20W class laser at Paranal, showing the high variability of the return flux versus the pointing direction.

\begin{figure}[H]
    \centering
    \includegraphics[width=1.0\columnwidth]{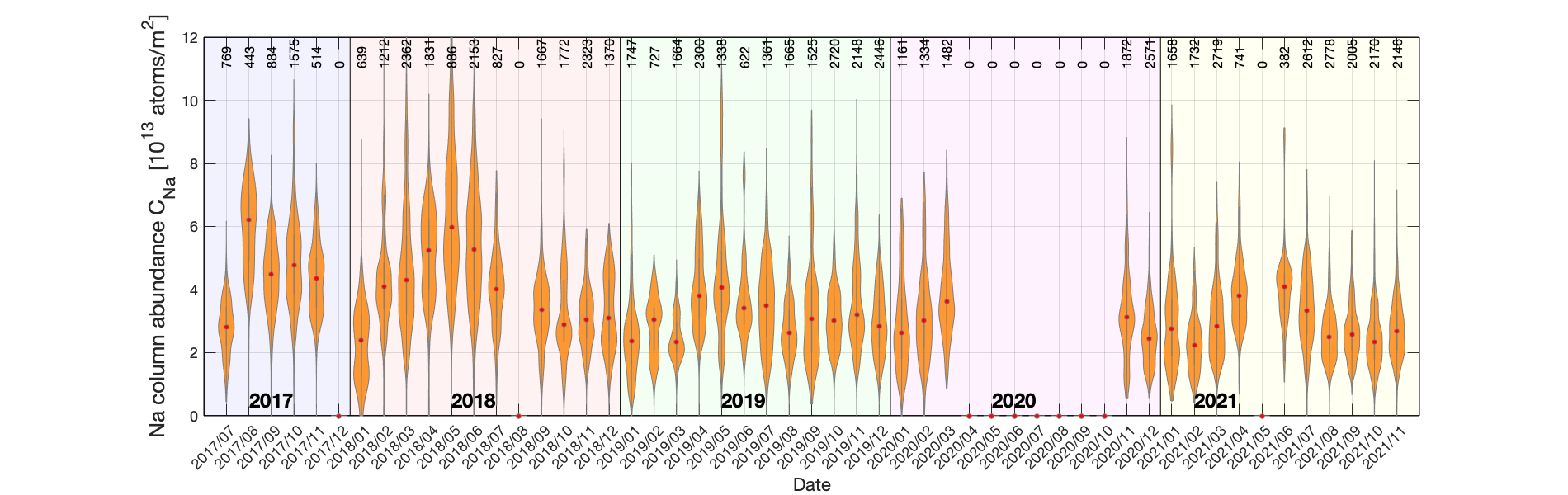}
    \caption{Statistics of Na column abundance per month.}
    \label{fig:NaAbundanceVsTime-GALACSI-Violin}
\end{figure}

The second parameter affecting the LGS flux return is the Na column abundance, which varies with time, on both long- and short-time scales. The Na column abundance in atoms/m$^2$ can be retrieved from the return flux measured on GALACSI, by correcting for the effect of the pointing direction. Fig.\ref{fig:NaAbundanceVsTime-GALACSI-Violin} shows the statistics of the Na column abundance per month. The “violin” representation shows the distribution of the Na column abundance from all the measurements of each month. The median value is shown as a red dot. The months when too few measurements were obtained to be statistically relevant have been marked with a 0. The known seasonal cyclic variation between the summer (low return) and winter (high return) months can be seen on the measurements. Differences can also be seen from year to year in the global average. Atmosphere studies show that the Na abundance is \textcolor{black}{correlated} to the Sun activity cycle of 11 years\cite{1979JGR....84.1543S}. The last peak of the Sun activity was in 2014, with a decrease since that date. The prediction for the Sun activity show a minimum in 2021 and the start of a new cycle with the next peak in 2025.

The morphology and temporal variations of the Na layer has been studied with different methods and at different sites\cite{2014A&A...565A.102P,CastroAO4ELT5,Neichel2013}. No measurements \textcolor{black}{(morphology and/or temporal)} of this type exist for the VLT or ELT, but the three studies reported similar results for very different location on Earth and their results can thus be considered representative for the ELT also. The following parameters can thus be used to describe the general structure of the Na layer, with all altitude given above sea level: mean centroid altitude of 90.8$\pm$0.1km, altitude range (containing 95$\%$ of the photons) of 13.1$\pm$0.3km (minimum width of 6 km a maximum one of 21 km), mean lower edge altitude of 81.7$\pm$0.1km, mean upper edge altitude of 104.9$\pm$0.3km. 

The mean altitude also shows variations \textcolor{black}{during the course of a night}. The power spectral density of these variations can be represented by a power law: $P_a(\nu)=\alpha\nu^{\beta}$, with $\nu$ the fluctuations frequency. The amplitude $\alpha$ and power coefficient $\beta$ show variations from night to night, but can be represented by the following mean values: $\alpha=34.4^{+5.6}_{-4.8} m^2 Hz^{-1}$ and $\beta=-1.87\pm0.02$.

The Na density vertical structure is also showing a high variability, and it is considered difficult to define any kind of typical vertical profile. An attempt has still been made at classifying the zoo of profiles measured on-sky into seven different classes\cite{2014A&A...565A.102P}.

\subsection{Toward LGSWFS for the ELT}
In this section, and based on the Sodium inputs derived above, we evaluate the boundary conditions recommended for the main LGSWFS parameters for the ELT, which are the number of sub-apertures (pupil sampling), the field-of-view per sub-aperture and the pixel scale. \textcolor{black}{The intention here is not to explore all the potential combinations of these parameters, but rather to provide what we think are boundary conditions to provide a LGSWFS design that would be robust to changing conditions. There may be ways to push each of these boundaries by implementing custom hardware, or smart software solutions, and we are not claiming that the proposed boundaries are hard limits. We only highlight one potential implementation, that we know will be working.}
Of course, hidden behind this trade-off are the detectors characteristics, which are the number of pixels, the frame rate, and the Read-Out Noise (RON). Depending on the detector availability and performance, one can adjust the LGSWFS parameters to fit into a given detector format, and the sensitivity study presented here allows that. Here we have provided boundaries for the LGSWFS parameters which allow to optimize the performance, assuming that suitable detectors are available (e.g. Ke, Pedreros et al.\cite{10.1117/1.JATIS.8.2.021511}). These trade-offs are summarized in Table \ref{tbl:tablechoice}. Basically, and based on the trade-off between pupil sampling, LGS spot truncation and LGS spot sampling, the recommendation for the ELT would be to deploy LGSWFSs with $>$64$\times$64 subapertures, $>$16" FoV and a pixel scale $<$1.15". 
The overall error budget would increase if the parameters of the LGSWFSs deviate from the ones indicated above and if no specific hardware or software solutions are implemented.
Note that a practical implementation based on these numbers is described in Ke, Pedreros et al.\cite{10.1117/1.JATIS.8.2.021511}.



\begin{table}[H]
\centering
\caption{LGSWFS main parameters for an ELT (40m) telescope}
\begin{tabular}{ccp{12cm}} 
\hline
   Parameters & range  & Comments \\ 
  \hline
subaps & $>$64$^2$ & See Fig.~\ref{fig:super_res_vs_sspup}. Use of super resolution allows to use less subapertures than DM actuators, but the larger the sampling, the better the performance. Going down to 64$\times$64 has negligible  impact on the performance. \\
FoV & $>$16$^2$ arcsec$^2$ & See Fig.~\ref{fig:MAORYonaxisHOerrorWFSfov}. The larger the FoV, the better the performance. Going above 16arcsec allows to minimize the truncation error for the worst Na profiles\\
Pixel Scale & $<$1.15arcsec & This is set by the minimum expected LGS spot size, and to maintain a good sampling and a decent niose propagation (see Fig. \ref{fig:noise}) . \\
RON & $<$3 $e-$ & with a good target around 2.5 $e-$. This value is fixed to be almost photon noise limited in the typical range of LGS flux return (see Fig. \ref{fig:noise})  \\
  \end{tabular}
  \label{tbl:tablechoice}
\end{table}


\section{Conclusion}
In this paper we have explored the key LGSWFS parameters impacting the performance when deployed at the ELT scale. When going from the current generation of 8-10m telescopes to a 40m scale, the first challenge comes from the LGS spot elongation. Where it only represents a few arcseconds for 8-10m telescope, it becomes a major limitation for ELT, with expected elongations of up to 25arcsec. Designing an LGSWFS then becomes a challenge, as one has to manage both the FoV and the small axis sampling. As this is out of reach of current detector technologies, a trade-off has to be made. In this paper, we open a new dimension in this trade-off, but making use of super resolution, and by taking advantage of the fact that we have several LGSWFS looking at an almost similar turbulence volume. Combining the signal from the multiple WFSs, we show that we can go beyond the historical paradigm of matching the subapertures count with the DM actuators geometry. Thanks to super resolution, we show that fewer subapertures per WFS can be used, which opens the path 
\textcolor{black}{for LGSWFS design based on existing or realistic detector hardware.}

\acknowledgments 
This work  benefited from the support the following grants: WOLF ANR-18-CE31-0018, F-CELT ANR-21-ESRE-0008 and LabEx FOCUS ANR-11-LABX-0013 of the French National Research Agency (ANR), OPTICON H2020 (2017-2020) Work Package 1 (Calibration and test tools for AO assisted E-ELT instruments) grant number 730890, Action Spécifique Haute Résolution Angulaire (ASHRA) of CNRS/INSU co-funded by CNES.


\bibliography{references}   
\bibliographystyle{spiejour}   

\end{document}